\documentclass[11pt]{article}
\usepackage{moriond,epsfig}
\usepackage{amsmath,amssymb}
\newcommand{\Mat}[1]{{{\boldsymbol{#1}}}}
\newcommand{\abs}[1]{\left\vert#1\right\vert}

\def\be{\begin{equation}}
\def\ee{\end{equation}}
\def\bea{\begin{eqnarray}}
\def\eea{\end{eqnarray}}

\begin{document}
\vspace*{4cm}
\title{POINT-PARTICLE LIMIT IN A SCALAR THEORY OF GRAVITATION \\AND THE WEAK EQUIVALENCE PRINCIPLE}

\author{ Mayeul ARMINJON }

\address{Laboratoire ``Sols, Solides, Structures'' [Unit\'e Mixte de Recherche of the CNRS], 
\\ BP 53, F-38041 Grenoble cedex 9, France}

\maketitle\abstracts{
A scalar theory with a preferred reference frame is summarized. To test that theory in celestial mechanics, an ``asymptotic" post-Newtonian (PN) scheme has been developed. This associates a conceptual family of self-gravitating systems with the given system, in order to have a true small parameter available. The resulting equations for a weakly-self-gravitating system of extended bodies include internal-structure effects. The internal-structure influence subsists at the point-particle limit---a violation of the weak equivalence principle. If one could develop an ``asymptotic" approximation scheme in general relativity also, this could plausibly be found there also, in a gauge where the PN space metric would not be ``conformally Euclidean".
}

\section{Introduction}
The weak equivalence principle (WEP), which states that the acceleration due to gravity is the same for any kind of matter, applies undoubtedly to Newton's gravity. This is because Newton's third law implies that the gravitational self-force over a body sums to zero---as it may be also checked on the continuous form of Newton's attraction law, based on the Poisson equation. It is taken for granted that the WEP applies also to general relativity (GR), because the interpretation of gravitation as a purely geometric phenomenon seems to make this obvious. For instance, it is well-known that, by applying the local ``conservation equation" for the energy-momentum tensor $\mathbf{T}$:
\begin{equation} \label{continuumGR}
T_{\mu;\nu}^{\nu} = 0				          
\end{equation}
to dust matter, one finds that each volume element of the dust continuum has a geodesic motion. (More generally, Eq.~(\ref{continuumGR}) seems to imply that matter flows universally in the curved space-time, independently of which kind of matter is present.) However, the space-time metric, that determines the geodesic lines, is certainly influenced by the presence of the very body whose motion is sought for, however small be its mass. And since this influence is nonlinear, it is not so obvious that it must cancel in the limit where the size of the body evanesces. Hence the motion of even a small dust object might after all depend, perhaps, on the structure of this object, {\it e.g.} it might depend on its density field and its internal motion. \\

The work \cite{A33} summarized here came as a natural follower of a study \cite{A32} on the equations of motion of the mass centers in a self-gravitating system of extended bodies in a scalar alternative theory of gravitation: \cite{A18,A15,A20} There, it was found that the internal structure of the bodies does play a role on the motion of their mass-centers in this theory. Hence, it appeared necessary to investigate the point-particle limit in this theory. Moreover, with the ``asymptotic" scheme of PN approximation (PNA) \cite{A23,B22} which has been used in the former work, \cite{A32} the internal-structure influence appears rather unavoidable, quite independently of the theory considered. \cite{A32} Therefore, in the present paper, we shall also briefly discuss what might happen in GR.

\section{Basic Principles of the ``Scalar Ether-Theory''} \label{principles}
This is a bimetric theory, thus it endows space-time with two Lorentzian metrics: a flat metric $\Mat{\gamma}^0$ (the ``background", or ``prior-geometrical" metric), and a curved metric $\Mat{\gamma}$ (the ``physical" metric, which is more directly related with physical space and time measurements). It admits a preferred reference frame $\mathrm{E}$ (``ether"). Most equations are written in $\mathrm{E}$ and are merely space-covariant. Space-time is assumed to be the product $\mathbf{R}\times \mathrm{M}$, where $\mathrm{M}$ is the ``space" manifold, {\it i.e.} the set of the positions $\mathbf{x}$ in the preferred reference frame $\mathrm{E}$. Metric $\Mat{\gamma}$ is related to $\Mat{\gamma}^0$ through the scalar field $f$ as follows. We postulate a gravitational dilation of time standards, making measured time intervals smaller:
\begin{equation} \label {localtime}
		 			dt_\mathbf{x}/dT = \sqrt{f(\mathbf{x},T)} \equiv \beta(\mathbf{x},T)  \leq 1,
\end{equation}
where $t_\mathbf{x}$ is the ``local time" in $\mathrm{E}$, measured by a clock at a fixed point $\mathbf{x}$ in the frame $\mathrm{E}$, and where $T$ is the ``absolute time". The latter is the inertial time associated with $\Mat{\gamma}^0$ in the frame $\mathrm{E}$, it is also the time which would be measured at a point $\mathbf{x}$ bound to the frame $\mathrm{E}$ {\it and} far enough from any massive body. And we postulate a gravitational contraction of material objects (including length standards, thus making measured distances larger than distances evaluated with the Euclidean metric), only in the direction of the gravity acceleration: 
\begin{equation} \label{contraction}	
   dl = \left\{ \begin{array}{cc}
dl^0 & \mathrm{if}\,\mathbf{dx}\equiv (dx^i) \,\mathrm{orthogonal}\,\mathrm{to}\,\mathbf{g},\,i.e.\, \Mat{g}^0(\mathbf{g},\mathbf{dx})=0;\\ 
dl^0/\sqrt{f(\mathbf{x},T)} & \mathrm{if} \, \mathbf{dx} \,\mathrm{parallel} \, \mathrm{to}\,\mathbf{g}. 
\end{array} \right. 
\end{equation}
Here $\Mat{g}^0$ is the Euclidean space metric associated with $\Mat{\gamma}^0$ in the frame $\mathrm{E}$ and $dl^0 \equiv (\Mat{g}^0(\mathbf{dx},\mathbf{dx}))^{1/2}$ is the corresponding line element, similarly $dl$ is the line element defined with $\Mat{g}$, where $\Mat{g}$ is the Riemannian space metric associated with $\Mat{\gamma}$, also in the frame $\mathrm{E}$; and the gravity acceleration is:
\begin{equation} \label {g_grav}
		 			\mathbf{g}=-\frac{c^2}{2}\nabla_0f, \quad \nabla_0f\equiv \mathrm{grad}_{\Mat{g}^0}f, \quad (\mathrm{grad}_{\Mat{g}^0}f)^i\equiv g^{0ij}f_{,j}.
\end{equation}
(Matrix $(g^{0ij})$ is the inverse of the component matrix $(g_{0ij})$ of the Euclidean space metric $\Mat{g}^0$.) In coordinates bound to the frame $\mathrm{E}$ (and such that $x^0=cT$), the components of metric $\Mat{\gamma}$ are: $\gamma_{00} = f, \,  \gamma_{ij} = - g_{ij},  \,  \gamma_{0i} = 0$, the $g_{ij}$ 's being deduced from (\ref{contraction}).  \cite{A15,A23} \\

Dynamics of a test particle is governed by a relativistic extension of Newton's second law:
\begin{equation} \label{Newtonlawmasspoint} 
\mathbf{F}_0 + m(v)\mathbf{g} = \frac{D\mathbf{P}}{Dt_\mathbf{x}},		 		 
\end{equation}
where $\mathbf{F}_0$ is the non-gravitational ({\it e.g.} electromagnetic) force, $\mathbf{v}\equiv d\mathbf{x}/dt_{\mathbf{x}}$ the velocity and $v\equiv \Mat{g}(\mathbf{v},\mathbf{v})^{1/2}$ its modulus; $m(v) \equiv m(0)\gamma_v$ is the relativistic mass  ($\gamma_v$ is the Lorentz factor); $\mathbf{P} \equiv m(v) \mathbf{v}$ is the momentum; and $D\mathbf{w}/Dt_{\mathbf{x}}$ is the appropriate time-derivative of a (spatial) vector $\mathbf{w}$ in the space $\mathrm{M}$ endowed with the time-dependent metric $\Mat{g}$. \cite{A15,A16} In the static case ($f_{,0}=0$), that dynamics implies Einstein's motion along geodesics of the curved space-time metric $\Mat{\gamma}$. \cite{A16}  For a dust continuum, we may apply (\ref{Newtonlawmasspoint}) pointwise and this implies \cite{A20} the following equation: 
\begin{equation} \label{continuum}
T_{\mu;\nu}^{\nu} = b_{\mu}				          
\end{equation}
(compare with Eq.~(\ref{continuumGR})), where $b_\mu$ is defined by
\begin{equation} \label{definition_b}
b_0(\mathbf{T}) \equiv \frac{1}{2}\,g_{jk,0}\,T^{jk}, \quad b_i(\mathbf{T}) \equiv -\frac{1}{2}\,g_{ik,0}\,T^{0k}. 
\end{equation}
(Indices are raised and lowered with metric $\Mat{\gamma}$, unless mentioned otherwise. Semicolon means covariant differentiation using the Christoffel connection associated with metric $\Mat{\gamma}$.) Universality of gravity means that the same equation must remain true for any material medium. {\it Thus, the dynamics of a test particle, as well as the dynamical field equation for a continuous medium, fully obey the ``weak equivalence principle", just as they do in GR.}\\

In the ``non-cosmological case'', the equation for the scalar field $f$  is
\begin{equation} \label{field}
\Delta f- \frac{1}{f} \left(\frac{f_{,0}}{f}\right)_{,0}= \frac{8\pi G}{c^2} \sigma \qquad (x^0 = cT),
\end{equation}
with $\Delta \equiv \mathrm{div}_{\Mat{g}^0} \mathrm{grad}_{\Mat{g}^0}$ the usual Laplace operator defined with the Euclidean metric $\Mat{g}^0$,  with $G$ Newton's gravitation constant, and with $\sigma \equiv (T^{00})_{\mathrm{E}}$ the mass-energy density in the ether frame, where $\mathbf{T}$ is the ``mass tensor'', {\it i.e.} the energy-momentum tensor in mass units. 
\footnote{In the most general case, applicable to a heterogeneous universe on a cosmological time-scale, the gravitational field is not directly the field $f$ but instead the field of the ``ether pressure'' $p_e$, and a different field equation applies. But this reduces to (\ref{field}) if the time variation of the ``reference pressure" $p_e^\infty (T)\equiv \mathrm{Sup}_{\mathbf{x} \in\mathrm{M}} p_e(\mathbf{x},T)$ is neglected. \cite{A15} \\The analysis of homogeneous cosmological models \cite {A28} shows that the latter variation takes place over long time scales indeed, of the order of $10^8$ years. 
}

\section{Asymptotic PN approximation for relativistic theories of gravitation}\label{PNA}
In order to obtain asymptotic expansions of the fields as function of a small field-strength parameter $\lambda$, one should be able to associate a (conceptual) one-parameter {\it family} $(\mathrm{S}_\lambda)$ of gravitating systems with the physically given system $\mathrm{S}$. This means that one should be able to define a family of boundary-value problems. In relativistic theories of gravity, the natural boundary-value problem is the initial-value problem. This applies \cite {A23} to the investigated scalar theory. Thus one should define a family of initial-value problems. In GR, an approximation scheme based on a such family of initial conditions has been introduced by Futamase \& Schutz,\cite{FutaSchutz} although their initial condition for the space metric was very special. They derived (not detailed) expansions of the local equations but no ``global" equations {\it i.e.} for the mass centers of an N-body system. In the present scalar theory, detailed local equations, \cite{A23} and global ones, \cite{A25-26} were derived in a general case. \\

When trying to define a family suitable for the Newtonian limit (and indeed including a precise definition of the latter), one is led to observe that there is an exact similarity transformation in Newtonian gravity for a perfect-fluid system: \cite{A23,FutaSchutz} If $p^{(1)}$ (pressure), $\rho^{(1)}=F^{(1)}(p^{(1)})$ (density), $\mathbf{u}^{(1)}$ (velocity), and $U_{N} ^{(1)}$ (Newtonian potential), obey the Euler-Newton equations, then, for any
$\lambda>0$, the fields 
\begin{equation}\label{similarity1}
p^{(\lambda)}(\mathbf{x},T)=\lambda^2p^{(1)}(\mathbf{x},\sqrt{\lambda}\ T), \quad
\rho^{(\lambda)}(\mathbf{x},T)=\lambda\rho^{(1)}(\mathbf{x},\sqrt{\lambda}\ T),
\end{equation}
\begin{equation}\label{similarity2}
U_{N}^{(\lambda)}(\mathbf{x},T)=\lambda U_{N}^{(1)}(\mathbf{x},\sqrt{\lambda}\ T), \quad
\mathbf{u}^{(\lambda)}(\mathbf{x},T)=\sqrt{\lambda}\ \mathbf{u}^{(1)}(\mathbf{x},\sqrt{\lambda}\ T),
\end {equation}
also obey these equations---provided the state equation for system $\mathrm{S}_\lambda$ is 
$F^{(\lambda)}(p^{(\lambda)}) = \lambda F^{(1)}(\lambda^{-2} p^{(\lambda)})$.  To define the Newtonian limit in the scalar theory, we just apply the similarity transformation (\ref{similarity1})-(\ref{similarity2}) to the initial data for the gravitating system of interest (with suitable modifications, in particular $V \equiv (c^2/2)(1-f)$ is substituted for $U_N$). \cite{A23} This defines a system $\mathrm{S}_\lambda$ as the solution of this initial-value problem $\mathrm{P}_\lambda$, and one expects that, as $\lambda \rightarrow 0$, the solution fields admit expansions in powers of $\lambda$, whose dominant terms have the same orders in $\lambda$ as in Eqs.~(\ref{similarity1})-(\ref{similarity2}). Using these expansions for the system of interest, $\mathrm{S}$, is then justified insofar as the value $\lambda_0$ of $\lambda$ for $\mathrm{S}$ is small enough. It is then easy to check that, by adopting $[\mathrm{M}]_\lambda = \lambda[\mathrm{M}]$ and $[\mathrm{T}]_\lambda = [\mathrm{T}]/\sqrt{\lambda}$  as the new units for the system $\mathrm{S}_\lambda$ (where $[\mathrm{M}]$ and $[\mathrm{T}]$ are the starting units of mass and time), all fields become order $\lambda^0$, and the small parameter $\lambda$ is proportional to $1/c^2$ (indeed $\lambda=(c_0/c)^2$, where $c_0$ is the velocity of light in the starting units). Hence, the derivation of PN expansions is easy: {\it all fields} depend on the small parameter $1/c^2$, and are thus expanded in powers of $1/c^2$, each with $n+1$ terms ($k = 0, ..., n$). Thus, each exact equation splits into $n+1$ {\it exact} equations (it is just coefficient identification for a polynomial in $\lambda$). The scalar theory admits consistent first-order expansions in $\lambda$ (or $1/c^2$). The first term ($k = 0$) is Newtonian gravity, hence there is a correct Newtonian limit. The $n=1$ approximation is the first PNA, that is linear in the PN fields (which correspond to $k=1$). We call this approximation scheme \cite{A23,B22} the ``asymptotic PNA", because it is a rigorous application of the usual method of asymptotic expansions for a system of partial differential equations.\\

In the standard PNA, developed for GR by Fock \cite{Fock59} and Chandrasekhar, \cite{Chandra65} $1/c^2$ is formally considered as small parameter, and the matter fields $p$, $\rho$ and $\mathbf{u}$ are not expanded. This implies, in particular, that the equations of the standard PNA are not exact ones resulting from a ``splitting'' of the starting equations into separate equations for the successive orders---as this is the case for the ``asymptotic'' PNA. In contrast with the latter, the standard PNA does not pertain to the usual method of asymptotic expansions, and the equations are different in the two schemes. \cite{A23,B22} However, it does not appear easy to build a general asymptotic PNA in GR, for in GR the initial conditions must verify the nonlinear ``constraint equations". It might be difficult (if it is at all possible) to build a family of initial conditions that would both satisfy the constraint equations and be suitable for the Newtonian limit in, say, a perfect-fluid system.

\section{PN equations of motion of the mass centers in the investigated theory}
The mass centers are defined as local barycenters of the {\it rest-mass density}, $\rho_\mathrm{exact}$, 
which expands as  
\be\label{expans_rho}
\rho_\mathrm{exact} = \rho_{(1)} + O(1/c^4), \qquad \rho_{(1)} \equiv \rho  + \rho_1/c^2. 
\ee
To get the 1PN equations of motion of the mass centers, the local 1PN equations of motion               (which are just the first-order expansion, in $\lambda \propto c^{-2}$, of the continuum dynamics equation (\ref{continuum}) as applied to a perfect fluid), are integrated inside the bodies. Integrating the zero-order and first-order expanded equations of the time component of the local equation of motion, in domain $\mathrm{D}_a$ occupied by body $(a) (a=1,...,N)$,	 leads to define the 1PN mass of body $(a)$ as

\begin{equation}\label{defPNmass}
  M_a^{(1)} \equiv \int_{\mathrm{D}_a}\rho_{(1)} d\mathsf{V} = M_a+M_a^1/c^2,\qquad M_a\equiv\int_{\mathrm{D}_a}\rho
  d\mathsf{V},\qquad M_a^1\equiv\int_{\mathrm{D}_a}\rho_1 d\mathsf{V},
\end{equation}
($\mathsf{V}$ is the Euclidean volume measure on the space $\mathrm{M}$), and this integration gives just $M_a=$ Const., $M_a^1 =$ Const.	 Thus, the 1PN mass center $\mathbf{a}_{(1)}$ of body $(a)$ is defined by
 \begin{equation}\label{defPNmasscent}
  M_a^{(1)}\mathbf{a}_{(1)}\equiv\int_{\mathrm{D}_a}\rho_{(1)}\mathbf{x}d\mathsf{V}=
  M_a\mathbf{a}+M_a^{1}\mathbf{a}_{1}/c^2,
\end{equation}
with
\begin{equation}\label{defmasscent-ord0-ord1}
  M_a\mathbf{a} \equiv \int_{\mathrm{D}_a}\rho\mathbf{x}d\mathsf{V},\qquad M_a^{1}\mathbf{a}_{1} \equiv \int_{\mathrm{D}_a}\rho_{1}\mathbf{x}d\mathsf{V}.
\end{equation}
Integrating the space components of the local PN equations of motion gives the sought equations of motion of the PN mass center $\mathbf{a}_{(1)}$. The 0-order equation is the Newtonian equation:
\begin{equation}\label{masscent-ord0}
  M_a\ddot{a}^i= \int_{\mathrm{D}_a} \rho U^{(a)}_{,i} d\mathsf{\mathsf{V}}
\end{equation}

($U$ is the zero-order coefficient in the expansion of $V \equiv (c^2/2)(1-f)$, it obeys the Poisson equation with the zero-order density $\rho$; thus it is the Newtonian potential; the superscript $^{(a)}$ designates the external part of $U$.) The PN correction (order 1 equation) has the form
\begin{equation}\label{masscent-ord1}
  M_a^1\,\ddot{a}_1^i = \int_{\mathrm{D}_a} f^1_i d\mathsf{V}.
\end{equation}
The ``PN force" density  $f^1_i$  depends in particular on the zero-order fields: 0-order pressure $p$, density $\rho$, velocity $\mathbf{u}$, Newtonian potential $U$. As a consequence, the internal structure of the bodies influences the motion already from the first PNA. \cite{A25-26,A32,A33} This follows naturally from using the ``asymptotic" method of PNA and, in the author's opinion, should hold true for GR if a (general) asymptotic PNA could be used in GR (see the end of Sect.~\ref{PNA}). 

\section{Point-Particle Limit. Reason for Conflict with the Weak Equivalence Principle}
We consider \cite{A33} the situation in which the size of one of the $N$ bodies, say (1), is a small parameter $\xi$. (This body could be an asteroid or a spacecraft.) Thus we consider a family $(\mathrm{S}'_\xi)$ of 1PN systems. (Since the equations of the asymptotic 1PN approximation make a closed exact system, we may content ourselves with those equations and hence forget the weak-field parameter $\lambda$---provided the system considered is indeed suitable for the PNA.) This family is defined by an initial data which is independent of $\xi$ apart from the size of the small body. The work consists then essentially in expanding the integrals on the r.h.s. of Eqs.~(\ref{masscent-ord0}) and (\ref{masscent-ord1}) as $\xi \rightarrow 0$, for the small body, {\it i.e.} $a=1$. The principal part of the Newtonian acceleration is quickly got from (\ref{defmasscent-ord0-ord1})$_1$ and (\ref{masscent-ord0}): it is
\be\label{accel_0_pplim}
\ddot{\mathbf{a}} = \nabla_0 U^{(1)}(\mathbf{a}) + O\left(\xi^2\right).
\ee
This means that the Newtonian point-particle limit is unproblematic, as expected. To expand the PN correction to the acceleration, we use the simplifying assumption that the zero-order (Newtonian) motion of the small body is a {\it rigid motion}. 

Then it remains merely a (rather lengthy) calculus. \cite{A33} We find that, at the point-particle limit: $\xi \rightarrow 0$, there remains a {\it structure-dependent part,} $\mathbf{A_S} = \mathrm{ord}(\xi^0)$, in the PN acceleration $\mathbf{A}_{(1)}$ of the small body. Moreover, in the case that there is just one massive body, assumed static and spherical, the limit of $\mathbf{A}_{(1)}$ differs from the PN acceleration of a test particle just by $\mathbf{A_S}$. This is a patent violation of the weak equivalence principle! In the general case, the structure-dependent acceleration $\mathbf{A_S}$ depends on a symmetric tensor $\mathbf{S}$, with
\be \label{def_S}
S_{kj} \equiv  \int_{D_1} \rho u_{1,k,j} \, d\mathsf{V}/M_1
\ee
(where $u_1$ is the Newtonian self-potential of body (1)), and it may be written in the form
\be \label{A_S}
\mathbf{A_S} = A_{\mathbf{S} \,\mathrm{max}} \mathbf{w},    
\ee 
with $\abs{\mathbf{w}}$ strongly orientation-dependent, and
\be \label{A_Smax}
A_{\mathbf{S} \,\mathrm{max}} \cong 2\times 10^{-7}d\quad  \mathrm{m/s}^2
\ee
on Earth ($d$ is an average density of the small body, expressed in g/cm$^3$). This magnitude seems dangerous, but it may not kill the theory. \cite{A33}\\

The {\it general reason} that makes this violation {\it a priori} possible is simply the fact that   the small body is gravitationally active and the theory is nonlinear! Since the PN part of the coordinate acceleration field, $\mathbf{\dot{u}}_1$, depends nonlinearly on (all) matter fields, we cannot even define which part of $\mathbf{\dot{u}}_1(\mathbf{x})$ comes from the body itself where $\mathbf{x}$ belongs, and which part comes from the other bodies. This nonlinear dependence is then carried over to the PN correction to the acceleration of the mass center, through Eq. (\ref{defmasscent-ord0-ord1})$_2$. Thus, there is no {\it a priori} reason why the acceleration of the mass center should depend only on the other bodies in a relativistic theory of gravitation, even if the body is very small. Now the {\it specific reason} that makes this violation actually occur is the presence in the PN spatial metric of derivatives $U_{,i}$  of the Newtonian potential. Indeed, the spatial Christoffel symbols depend therefore on the {\it second} derivatives $U_{,i,j}$, whose self-part is independent on $\xi$, hence survives in the point-particle limit (see Eq.~(\ref{def_S})). But the $U_{,i}$ 's are also there in Schwarzschild's metric (this is the ``anisotropy" of the latter---by the way, it is the exact solution of this scalar theory for the static spherically symmetric case \cite{A18,B22}). Hence, although the standard PN spatial metric of GR is conformally Euclidean,\cite{Fock59,Chandra65} and thus no $U_{,i}$ 's are there, we believe that the same violation could occur in GR, if one would (could) use an asymptotic PNA, and this in a gauge where the PN metric would be ``Schwarzschild-like" instead. \cite{A33}


\section*{References}

\begin{thebibliography}{99}

\bibitem{A33}
M. Arminjon, ``Equations of motion of the mass centers in a scalar theory of gravitation: The point particle limit", submitted, gr-qc/0301031.

\bibitem{A32}
M. Arminjon, ``Equations of motion of the mass centers in a scalar theory of gravitation: Expansion in the separation parameter", submitted, gr-qc/0202029.

\bibitem{A18}
M. Arminjon, {\it Rev. Roumaine Sci. Tech. -- M\'ec. Appl.} {\bf
42}, 27--57 (1997). Online at \\ 
\verb+http://geo.hmg.inpg.fr/arminjon/pub_list.html#A18+

\bibitem{A15}
M. Arminjon, {\it Arch. Mech.} {\bf 48}, 25--52 (1996). Online at\\
\verb+http://geo.hmg.inpg.fr/arminjon/pub_list.html#A15+

\bibitem{A20}
M. Arminjon, {\it Anal. Univ. Bucuresti -- Fizica} {\bf
47}, 3--21 (1998), physics/9911025.

\bibitem{A23}
M. Arminjon, {\it Roman. J. Phys.} {\bf 45}, 389--414 (2000), gr-qc/0003066.

\bibitem{B22}
M. Arminjon, in {\it Eighth Conf. ``Physical Interpretations of
Relativity Theory'', Proceedings} (M.C. Duffy, ed.,
Univ. of Sunderland/Brit. Soc. Philos. Sci., to appear) gr-qc/0305078.

\bibitem{A16}
M. Arminjon, {\it Arch. Mech.} {\bf 48}, 551--576 (1996). Online at \\ 
\verb+http://geo.hmg.inpg.fr/arminjon/pub_list.html#A16+

\bibitem{A28}
M. Arminjon, {\it Phys. Essays} {\bf 14}, 10--32 (2001), gr-qc/9911057.

\bibitem{FutaSchutz}
T. Futamase and B. F. Schutz, {\it Phys. Rev.} {\bf D28},
2363--2372 (1983).

\bibitem{A25-26}
M. Arminjon, {\it Roman. J. Phys.} {\bf 45}, 645--658 and 659--678
(2000), astro-ph/0006093
.

\bibitem{Fock59}
V. A. Fock, {\it The Theory of Space, Time and Gravitation} (1st
English edn., Pergamon, Oxford, 1959). (Russian original edn. 1955.)

\bibitem{Chandra65}
S. Chandrasekhar, {\it Astrophys. J.} {\bf 142}, 1488--1512 (1965).


\end{thebibliography}
\end{document}